# AI-Enabled Ultra-Low-Dose CT Reconstruction

Weiwen Wu[1], Chuang Niu[1], Shadi Ebrahimian[2], Hengyong Yu[3], Mannu Kalra[2], Ge Wang[1]

[1]Biomedical Imaging Center, Center for Biotechnology and Interdisciplinary Studies, Department of Biomedical Engineering, Rensselaer Polytechnic Institute, Troy, NY, USA

[2]Department of Radiology, Massachusetts General Hospital, Harvard Medical School, Boston, MA, USA

[3]Department of Electrical & Computer Engineering, University of Massachusetts Lowell, Lowell, MA, USA.

**Abstract:** By the ALARA (As Low As Reasonably Achievable) principle, ultra-low-dose CT reconstruction is a holy grail to minimize cancer risks and genetic damages, especially for children. With the development of medical CT technologies, the iterative algorithms are widely used to reconstruct decent CT images from a low-dose scan. Recently, artificial intelligence (AI) techniques have shown a great promise in further reducing CT radiation dose to the next level. In this paper, we demonstrate that AI-powered CT reconstruction offers diagnostic image quality at an ultra-low-dose level comparable to that of radiography. Specifically, here we develop a Split Unrolled Grid-like Alternative Reconstruction (SUGAR) network, in which deep learning, physical modeling and image prior are integrated. The reconstruction results from clinical datasets show that excellent images can be reconstructed using SUGAR from 36 projections. This approach has a potential to change future healthcare.

## I. Introduction

Chest CT is one of the most commonly performed imaging modalities only second to chest radiography. There is several-fold difference between radiation doses associated with chest CT as compared to chest radiography. Up until recently, use of chest CT in the United States was limited to symptomatic patients or those with known or suspected diseases. Since the conclusion of the National Lung cancer Screening Trial (NLST), use of chest CT has been extended to screening of asymptomatic patients who are at risk of lung cancer. The NLST demonstrated that annual screening of at-risk patients with CT is associated with 20% relative risk reduction of death from lung cancer as compared to chest radiography [1, 2]. To reduce potential risk associated with radiation dose from annual CT, a low-dose CT (LDCT) is recommended for lung cancer screening. However, the recommended target of 1.5 mSv for LDCT in average-size adult patients is still an order of magnitude higher than 0.1 mSv dose from two-projection (posteroanterior and lateral projections) chest radiographs [3].

Despite evidence for reduction of CT radiation dose by several folds as compared to the current standard of care, low-dose and ultra-low dose CT protocols are often not used [4-8]. For lung cancer screening LDCT, a recent study reported that nearly two-third of US scanner sites had median radiation doses above the recommended American College of Radiology (ACR) guidelines. Like lung nodules, kidney stones are also amenable for evaluation at lower radiation dose, and yet only 2% of the US sites in 2013 and 8% of the US sites in 2016 applied reduced dose CT protocols for assessing patients with renal colic [7, 8]. This trend remains despite the fact the ACR Appropriateness Criteria™ recommend for use of low dose CT in patients with suspected renal calculi as primary indication for their scanning [9]. The hesitation of adopting lower radiation doses may stem from both concern over loss of diagnostic information and lack of faith in existent dose reduction technologies and image reconstruction algorithms. The limitations of current dose reduction techniques and their adoption calls for better options and further improvements in dose reduction and image quality optimization.

Over the past years, many compressed sensing methods were developed to reduce radiation dose, such as [10]. As well known now, although the compressed sensing methods produce good image quality from low-dose CT data, deep learning methods demonstrate superior performance even from ultra-low-dose CT (ULD-CT) data. In other words, deep learning methods do bring new insight and additional power into ultra-low-dose CT reconstruction.

Conventional iterative reconstruction methods regularize a reconstructed image using prior knowledge. Often times, such an approach first embeds an intermediate reconstruction into a latent space and then refines the image aided by total variation [11, 12], low-rank representation [13], low-dimensional manifold [14], dictionary learning [15], and alike. These regularizers take advantages of formulated image properties, such as piece-wise constant or polynomial. In various reconstruction tasks, different regularized priors are used. Also, there are weighting/hyper parameters that must be empirically tuned. Finally, this class of reconstruction methods take a high computational cost, with ART-type [16] or SART/EM-type [17] iterations being involved. Although these methods can find good solutions under proper conditions; for example, Chambolle-Pock Primal-Dual TV minimization (CPPD-TV) works well for piece-wise constant phantoms [18], they would fail to address complicated reconstruction tasks from rather sparse data, such as reconstruction from a very small number of projections.

The physics-based models reflect laws of nature with embedded concepts of space-time, causality and manifolds [19]. Inverse problems with specific physical model occupy every field including science, engineering and medicine. Tomographic imaging, as one of typical inverse problems, achieves non-invasive visualization of internal structures of an opaque object. Over the past several years, deep learning algorithms have been rapidly developed for tomographic imaging, and achieved great results [20, 21]. Indeed, deep reconstruction can gain significantly over compressed sensing methods in terms of texture preservation and structural fidelity [22].

The deep learning algorithms were successfully developed for limited data reconstruction [23], which are classified into three categories: image-to-image, deep image-domain iteration, and end-to-end methods. First, an image-to-image reconstruction network maps a poor image to a ground truth counterpart after a big-data-driven training process; for example, the deep convolutional neural network (FBPConvNet) [24], convolutional neural network (CNN) [25], wavelet transform based U-net [26, 27], residual encoder-decoder CNN (RED-CNN) [23], densenet deconvolution network (DD-Net) [28], and others [29]. These methods obtained good results efficiently. However, the original measurement and synthesized data from the network output can be inconsistent, resulting in blurred details and image artifacts. Second, a trained network as a non-standard regularizer can be integrated into a classic iterative algorithm, referred to as a deep image-domain iterative framework. There are multiple examples, such as the momentum-Net [30] and deep BCD-net [31]. Although these networks achieved great metrics in reported cases, they are difficult to be used in practice due to a huge computational overhead with embedding ART-type [16] or SART/EM-type [17] iterations. Furthermore, the learned network is used on the reconstructed image, and it is restricted by the image quality produced using the iterative algorithm. Finally, end-to-end deep reconstruction methods directly transform tomographic measurement into images using a trained network, such as AUTOMAP [21]. Since AUTOMAP utilizes a fully convolutional architecture, it is difficult to implement in practice where the size of input data is too large to be handled on common computing platforms. To map Radon data to an image, the iRadonMap was proposed for CT [32]. Inspired by the information workflow of iterative reconstruction, un-rolling-based networks were proposed, including the learned experts' assessment-based reconstruction network (LEARN) [33], manifold and graph integrative convolutional network [34], analytical and iterative reconstruction network (AirNet) [35], Meta-Inv [36], $R^2$edSCAN [37], FISTA [38], etc. These unrolling-based networks obtained encouraging results for sparse-view reconstruction with high spatial resolution (<1mm in clinical scenario). However, they still are difficult to reconstruct high-quality images from ULD-CT data, such as 36 views along a limited angular range of about 150° (as shown in Figure 1), which will be addressed in this paper.

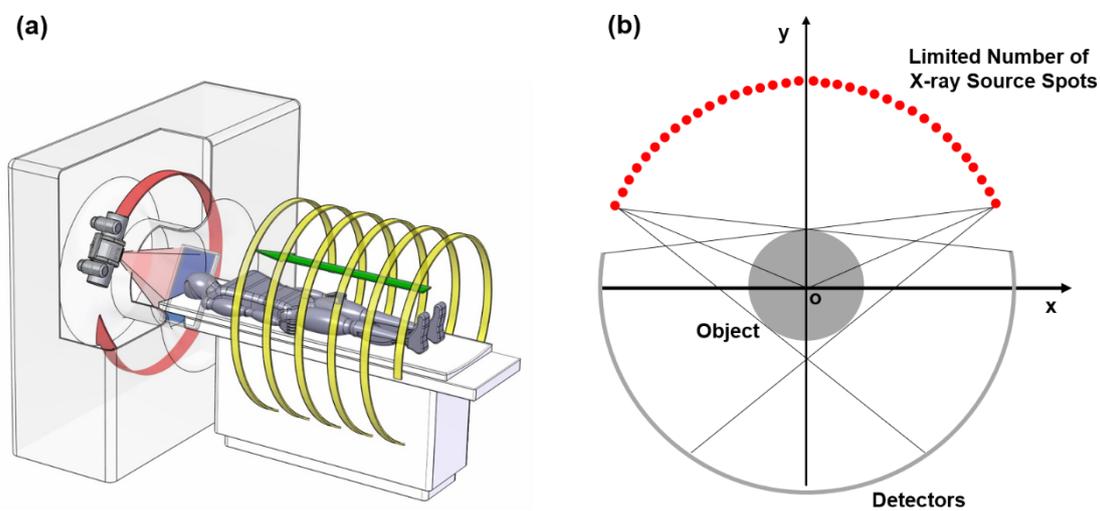

Figure 1. CT scan with data acquisition flexibilities (either full or sparse scans in helical or circular modes). (a) A typical CT scanner with a wide-pulsing window for selected data segments [39]; and (b) 36 distributed x-ray source spots over a limited angle range of about 151°.

To improve deep reconstruction quality from ultra-low-dose data, here we propose a Split Unrolled Grid-like Alternative Reconstruction network (SUGAR) to address such this rather challenging CT problem. Our results suggest a major improvement with this deep reconstruction approach over the compressed sensing based methods such as CPPD-TV [18]. Since the training and testing datasets come from clinical patient datasets, our results are clinically relevant. In addition, our SUGAR method still obtained better performance in comparison with other advanced reconstruction networks that were very recently developed. In the context of the extensive efforts over the past years in deep reconstruction, the unique contributions of this paper include (1) our SUGAR network is based on the split-Bregman strategy that represents a state of the art for numerical optimization; (2) the image reconstruction process is unrolled into a single network, where the sampling transform and the image recovery are implemented with multiple-scale convolution and deconvolution layers to capture image features using learned regularizing and coupling parameters; and (3) a novel two-stage reconstruction strategy to recover high quality images.

The rest of the paper is organized as follows. In the next section, we formulate an optimization-based reconstruction model and motivate our SUGAR network. Then, we describe the SUGAR network architecture and its features. In the third section, we report our representative results showing the feasibility of ultra-low-dose reconstruction enabled by the SUGAR method. It yields the best image quality from ultra-low-dose projections and is superior to the results using competing methods. In the last section, we discuss related issues and conclude the paper.

## II. Methods

### A. Linear Imaging Model

The few-view image reconstruction problem for x-ray CT is to recover an underlying image from sparse projection data based on the corresponding measurement model. Let $A \in \mathbb{R}^{m \times N} (m \ll N)$ be a discrete-to-discrete linear transform representing a CT system from image pixels to detector readings; $y \in \mathbb{R}^m$ is an original measurement, $e \in \mathbb{R}^m$ is data noise within $y$, and $x \in \mathbb{R}^N$ is the image to be reconstructed, and most relevantly $m \ll N$ signifies that the inverse problem is highly underdetermined. Furthermore, $H$ represents a sparsifying transform to enforce prior knowledge on the image. In this setting, the image reconstruction task with sparsity prior is expressed as follows:

$$x^* = \operatorname*{argmin}_{x} \ \|Hx\|_0, \quad \text{subject to } y = Ax + e, \qquad (1)$$

where $\|.\|_0$ represents the $\ell_0$-norm. Because Eq. (1) demands the $\ell_0$-norm optimization, it is an NP-hard problem [40]. However, it is feasible to relax the $\ell_0$-norm optimization in Eq. (1) to the $\ell_1$-norm surrogate. Then, Eq. (1) is relaxed to the following:

$$x^* = \operatorname*{argmin}_{x} \ \|Hx\|_1, \quad \text{subject to } y = Ax + e. \qquad (2)$$

In most cases of CT image reconstruction, the optimization problem Eq. (2) can be solved using an iterative algorithm. A desirable solution to Eq. (2) can be found in the set expended by $H$ with an image generating data close to $y$. In other words, the optimization problem of Eq. (2) is equivalent to the following unconstrained problem:

$$x^* = \underset{x}{\operatorname{argmin}} \frac{1}{2}\|y - Ax\|_2^2 + \lambda\|Hx\|_1, \tag{3}$$

where $\lambda > 0$ balances the data fidelity $\frac{1}{2}\|y - Ax\|_2^2$ and the image prior $\|Hx\|_1$. The goal of Eq. (3) is to find an optimized solution by minimizing the whole objective function. To solve this problem, the split-Bregman strategy can be employed. We first split the data fidelity and regularized prior by introducing $z$ to re-express Eq. (3):

$$\{x^*, z^*\} = \underset{\{x,z\}}{\operatorname{argmin}} \frac{1}{2}\|y - Ax\|_2^2 + \lambda\|Hz\|_1, \text{ subject to } z = x. \tag{4}$$

Then, the constrained optimization problem can be converted into an unconstrained optimization problem by introducing an error variable $f$ as follows:

$$\{x^*, z^*, f^*\} = \underset{\{x,z,f\}}{\operatorname{argmin}} \frac{1}{2}\|y - Ax\|_2^2 + \frac{\lambda_1}{2}\|x - z - f\|_2^2 + \lambda\|Hz\|_1. \tag{5}$$

There are three variables in Eq. (5), which can be handled by solving the following three sub-problems alternatively:

$$x^{(k+1)} = \underset{x}{\operatorname{argmin}} \frac{1}{2}\|y - Ax\|_2^2 + \frac{\lambda_1}{2}\|x - z^{(k)} - f^{(k)}\|_2^2, \tag{6}$$

$$z^{(k+1)} = \underset{z}{\operatorname{argmin}} \frac{\lambda_1}{2}\|x^{(k+1)} - z - f^{(k)}\|_2^2 + \lambda\|Hz\|_1, \tag{7}$$

$$f^{(k+1)} = \underset{f}{\operatorname{argmin}} \frac{\lambda_1}{2}\|x^{(k+1)} - z^{(k+1)} - f\|_2^2. \tag{8}$$

**Subproblem with $x$:** Eq. (6) can be solved by setting the derivative to zero:

$$A^T(Ax - y) + \lambda_1(x - z^{(k)} - f^{(k)}) = 0, \tag{9}$$

By adding $(A^TA + \lambda_1)x^{(k)}$ into both sides of Eq. (9) and simplifying it, we have the following equation

$$(A^TA + \lambda_1)x = (A^TA + \lambda_1)x^{(k)} - A^T(Ax^{(k)} - y) - \lambda_1(x^{(k)} - z^{(k)} - f^{(k)}). \tag{10}$$

Therefore, $x$ can be updated with

$$x^{(k+1)} = x^{(k)} - (A^TA + \lambda_1)^{-1}\left(A^T(Ax^{(k)} - y) + \lambda_1(x^{(k)} - z^{(k)} - f^{(k)})\right). \tag{11}$$

**Subproblem with $z$:** Eq. (7) can be solved via soft-thresholding:

$$z^{(k+1)} = H^* g_{\frac{2\lambda}{\lambda_1}}\left(H(x^{(k+1)} - f^{(k)})\right), \tag{12}$$

where $H^*$ represents the adjoint of $H$ satisfying $H^*H = I$ that is the identity transform, and the soft-thresholding kernel is defined as

$$g_{\frac{2\lambda}{\lambda_1}}(u) = \begin{cases} 0, & |u| < \frac{2\lambda}{\lambda_1} \\ u - sgn(u) \times \frac{2\lambda}{\lambda_1} & otherwise \end{cases}. \tag{13}$$

**Subproblem with $f$:** Eq. (8) can be iteratively updated as follows:

$$f^{(k+1)} = f^{(k)} - \eta(x^{(k+1)} - z^{(k+1)}), \tag{14}$$

where $\eta > 0$. In Eq. (5), there are three parameters that need to be empirically adjusted. Such an iterative reconstruction method takes a high computational cost to perform. More importantly, this iterative formulation does not take deep prior into account, which limits the reconstruction performance. Therefore, in our work we unroll this general iterative model into the feed-forward network so that we can train it in a data-driven fashion.

## B. Split Unrolled Grid-like Alternative Reconstruction (SUGAR) Network

### B.1 Network Architecture

In this section, we introduce our proposed interpretable neural network architecture, combining the split iterative reconstruction scheme and the unrolling strategy to solve the sparse-view CT image reconstruction problem. More precisely, the basic idea is to treat each iteration of the above iterative reconstruction scheme as a non-linear transform function $Q$ embedded in a neural network block. The overall architecture consists of multiple such deep blocks. Because each step in the network is well motivated, we can interpret its functionality accordingly, which is referred to as the Split Unrolling Grid-like Alternative Reconstruction (SUGAR) network for image reconstruction.

Up to now, most existing unrolled deep tomographic neural networks (including LEARN, *etc.*) target a low-dimension domain (256x256 pixels) instead of a high-dimension spatial domain (512x512 pixels). However, the low spatial resolution achieved by these unrolled tomographic networks may miss image details, leading to compromised imaging performance.

In this study, we designed a new strategy to effectively recover high-dimensional images from limited data, as shown in Figure 2(a). It can be observed there are two reconstruction steps: low-dimensional estimation (LE) and high-dimensional refinement (HR). In the LE step, we first achieve low-dimensional reconstruction with an LE network, where the underdeterminedness is under control in a data-driven manner. Then, the LE result is upsampled to a high-dimensional version with a HR network. Specifically, the whole network architecture for LE and HR is illustrated in Figure 2(b) and (c). Note that for optimal image reconstruction we need to both fully utilize the measurements and effectively minimize the sparsity of the resultant image.

In general, the reconstruction performance achieved by combining the image sparsity and the data consistency is still limited by lack of measurement data in the challenging cases of few-view tomography. Hence, it is necessary and powerful to use learnable nonlinear transforms to leverage data-driven prior and outperform compressed sensing inspired image reconstruction. In contrast to existing deep reconstruction networks, our approach based on well-designed neural blocks can optimize the imaging performance in reference to both the reconstructed image and the estimated error. Since the auxiliary error feedback variable reflects the information embedded in the residual image-domain, the proposed network architecture attempts to optimize image reconstruction in the image space with awareness of the residual error. Specifically, Figure 2(c) demonstrates the network-based reconstruction scheme consisting of the network forward transform (FT) $Q$ and the network backward transform (BT) $Q^*$. It can be seen that each transform contains multiple boxes, each of which includes a convolutional layer, a batch-norm (BN) layer and a rectified linear unit (ReLU) layer. The first convolutional layer consists of filters of 3×3 while the following convolutional network in FT also contains filters of 3 × 3. To deeply encode image features, the pooling layers are also used in the FT process. Such a design is not only beneficial to extract high-dimensional features but also effective to reduce the computational cost as compared to a fully convolutional layer. On the other hand, BT is an inverse of the feed forward transform. As shown in Figure 2(c), this network has a structure similar to FT except for the use of an unpooling layer instead of the pooling layer. It is pointed out that BT can convert the compressed feature maps back to an image satisfying that $Q^*Q(x) \approx x$. To facilitate the use of image features, the skip connections are made, as also shown in Figure 2(c). Hence, the whole network architecture makes it feasible to recover the target image from sparse/compressed measurements. Clearly, the optimization model of Eq. (3) can be expressed as

$$x^* = \underset{x}{\mathrm{argmin}} \frac{1}{2}\|y - Ax\|_2^2 + \lambda\|Qx\|_1. \tag{15}$$

Now, we need a deep learning method to solve this model. Specifically, we cast each iteration of the basic compressed sensing algorithm to a processing module. In this way, our network is interpretable in the compressed sensing perspective. That is, SUGAR attempts to update Eqs. (11), (12), and (14) by exploiting the power and flexibility of network-based transform functions. In essence, each iteration of SUGAR consists of an image reconstruction module (RM), deep estimation module (DM), and error correction module (EM), as illustrated in Figure 2(b). RM focuses on image reconstruction, DM estimates the residual error between the ground truth and a reconstructed image, and EM corrects the feedback error.

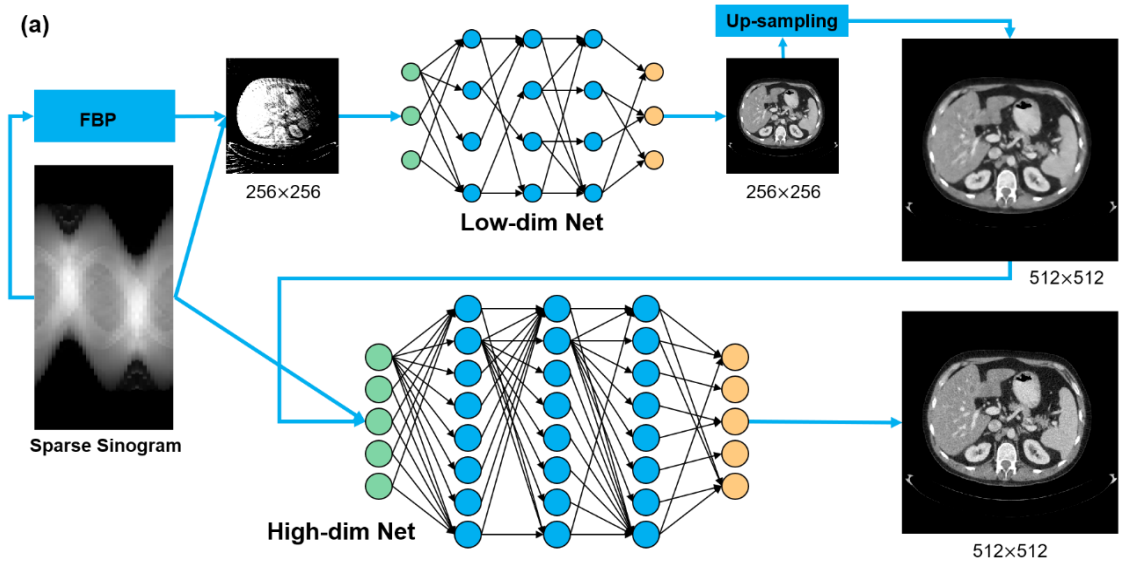

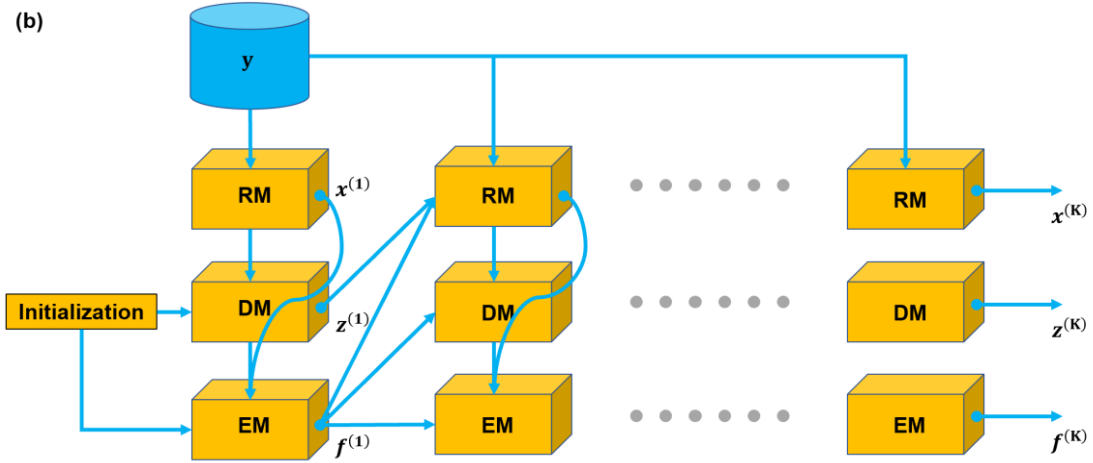

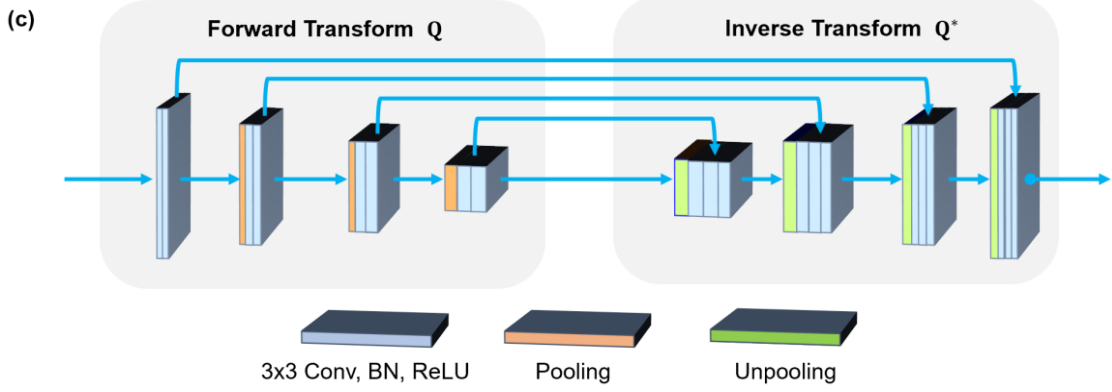

Figure 2. (a) The overall reconstruction strategy, (b) the unrolled iteration framework consisting of three modules RM, DM and EM, SUGAR network architecture for image reconstruction, deep estimation, and error correction respectively, and (c) a typical encoder-decoder network to model the compressed sensing inspired feed-forward and inverse transforms, respectively.

**B.2 RM Component**

The purpose of this module is to reconstruct an image according to Eq. (11). Taking the current iterates $x^{(k)}$, $z^{(k)}$ and $f^{(k)}$ as the input, we produce an updated image $x^{(k+1)}$. To improve the flexibility, we modify Eq. (11) into the following expression:

$$x^{(k+1)} = x^{(k)} - a\left(A^T(Ax^{(k)} - y)\right) - b(x^{(k)} - z^{(k)} - f^{(k)}), \qquad (16)$$

where $a$ and $b$ are two learnable parameters, which can be initially set to $1/\|A^T A + \lambda_1\|_2^2$ and $\lambda_1/\|A^T A + \lambda_1\|_2^2$, respectively. These parameters may vary with respect to the iteration index. If so, Eq. (16) can be expressed as

$$x^{(k+1)} = x^{(k)} - a^{(k)}\left(A^T(Ax^{(k)} - y)\right) - b^{(k)}(x^{(k)} - z^{(k)} - f^{(k)}). \qquad (17)$$

Note that $x^{(k)} - z^{(k)} - f^{(k)}$ is the coupling term via combination of all the outputs from the current iteration. The learnable parameters $a^{(k)}$ and $b^{(k)}$ are dynamically learnable as the iterative process goes on. In Eq. (17), the update to the reconstructed image can be treated as a gradient search step, which does not need any additional matrix inversion, with $A^T$ being conveniently approximated as FBP in this study.

**B.3 DM Component**

DM updates the variable $z$ that can be directly estimated via soft-thresholding, i.e.,

$$z^{(k+1)} = Q^* g_\epsilon\left(Q(x^{(k+1)} - f^{(k)})\right), \qquad (18)$$

where $\epsilon$ represents a soft-threshold satisfying with $\epsilon = 2\lambda/\lambda_1$. In an iterative reconstruction process, $\epsilon$ is a fixed constant. Eq. (18) can be decomposed into the three steps: image encoding, transform filtration, and image recovery. The encoding process of the variable $x^{(k+1)} - f^{(k)}$ is represented by the complicated nonlinear transform function $Q$ with the convolutional and rectified linear unit (ReLU) layers, i.e.,

$$z_1^{(k)} = Q(\mathbf{x}^{(k+1)} - f^{(k)}). \qquad (19)$$

Similar to the encoding process, the inverse network transform is performed on feature maps to recover a high-quality image, namely,

$$z^{(k+1)} = Q^*(z_1^{(k)}) = Q^*\left(Q(x^{(k+1)} - f^{(k)})\right). \qquad (20)$$

Comparing Eq. (18) with Eq. (20), it can be seen that the soft-thresholding operator has disappeared. In fact, the encoding-decoding process with the symmetric network-based transform functions can be viewed as an advanced version of soft-thresholding.

**B.4 EM Component**

As far as the error correction is concerned, with a dynamically adjusted updating rate $\eta$, Eq. (14) can be modified as follows:

$$f^{(k+1)} = f^{(k)} - \eta^{(k+1)}(x^{(k+1)} - z^{(k+1)}), \qquad (21)$$

where $\eta$ is a learnable network-specific and task-specific parameter.

**B.5 SUGAR Parameters**

Basically, the SUGAR network attempts to learn a set of parameters including the step-size $a^{(k)}$ and the coupling parameters $b^{(k)}$ in the RM component, the parameters of the network-based nonlinear transforms $Q^{(k)}$ and $Q^{*(k)}$ in the DM component, as well as the step length $\eta^{(k)}$ in the EM component. In brief, the proposed deep network can be described by the set of parameters taking the split iterative reconstruction scheme as a special case and outperforming it with data-driven adjustments to these parameters. In addition to the measurement data **y**, we also leverage the initialization of $\{x^{(0)}, z^{(0)}, f^{(0)}\}$. Regarding the loss function for network training, in this feasibility study we only used the peak signal-noise-ratio (PSNR) between the output and the ground truth.

**III. Clinical Results**

**A. Experimental Design**

**(1) Datasets and Comparison**

To validate the feasibility of ultra-low-dose CT imaging, we performed the clinical experiments on 2016 NIH-AAPM-Mayo Low-dose CT Grand Challenge datasets [22]. The datasets were obtained from Siemens Somatom Definition CT scanners at 120kVp and 200 mAs. Since the original scans are in helical cone-beam geometry, to handle such helical cone-beam measurements we sorted data into multiple slice fan-beam datasets. Specifically, we employed the single slice rebinning operation and took the flying focal spot into account [41]. The imaging parameters are as follows: the distances from x-ray source to detector and the system isocenter are 1085.6mm and 595mm, respectively. The curved cylindrical detector contains 736 units, each of which covers an area of 1.2858 × 1.0 mm$^2$, and there are 2304 views in a scan. Again, 946 projections are uniformly distributed over 151.875°. Then, 36 projections were extracted from the above 946 projections by selecting one per 28 projections to generate ultra-low-dose projections. The detector shift is 0.0013 radian. In this feasibility study, the size of a reconstructed image was set to 512×512 pixels, each of which covers 0.9×0.9 mm$^2$. Finally, a total number of 4,665 sinograms of 2,304×736 pixels were acquired from 10 patients at the normal dose setting, where 4,274 sinograms of 8 patients were employed for network training, and the rest 391 sinograms from the other 2 patients for network testing.

To highlight the advantages of deep learning reconstruction over compressed sensing inspired reconstruction, CPPD-TV [18] used in [10] was selected for comparison. The code of

CPPD-TV was provided by the authors in [42] and can be downloaded online (https://github.com/jakobsj/how_little_data). All parameters in CPPD-TV were optimized with the number of iterations set to 10,000. Next, representative deep learning based inverse imaging network LEARN [33] and FISTA [38] were selected. The experimental results demonstrate that our proposed network outperforms the advanced compressed sensing based method (i.e., CPPD-TV) as well as other unrolling-based deep reconstruction methods for sparse-view CT imaging. To quantify different reconstruction methods, PSNR was employed to measure the difference between reconstructed images and the reference. Here, the reconstructed image using FBP with full-scan projections was treated as the ground truth. Furthermore, the structural similarity (SSIM) index was used to compare between the reconstructed images and the reference.

In this study, all the source codes for deep learning reconstruction were programmed in Python with the Pytorch library on a NVIDA RTX3080. All programs were implemented on a PC (24 CPUs @3.70GHz, 32.0GB RAM) with Windows 10.

**(2) Network Training and Testing:** The Adam method was employed to optimize all of the networks [43]. To avoid the inconsistency in size between feature maps and the input, we padded zeros around the boundaries before convolution. The batch size for LE and HR in SUGAR was set to 1. The learning rate was decreased with the number of epochs. Here, the number of epochs was set to 40 for all the networks. The learning rate was set to $2.5 \times 10^{-4}$, and decreased by 0.8 after each of 5 epochs. In this study, the number of iterations for LE and HR networks were set to 70 and 30, respectively. In the testing process, 391 images were selected from two patients (L109, 291 slices; and L291, 100 slices).

**B. Performance Evaluation**
**(1) SUGAR Advantages in Terms of Feature Recovery**

Figure 3 shows the reconstructed results in three represent locations (Cases 1-3) from the patient L109, who has no lesion so that we only evaluate image quality in terms of anatomical features and details. Compared with the FBP results, CPPD-TV improved image quality with clear image features by incorporating the sparsity prior into reconstruction. However, CPPD-TV oversmoothed image details and edges. More importantly, the tissue features were missing. Also, since CPPD-TV is an iterative technique, it needs much longer time to obtain the final results. In this study, the total time of 24.18 minutes was taken for reconstructing an image from a sparse sinogram. Although the computational power is being improved, it is still a major challenge for use of such iterative methods for helical cone-beam CT reconstruction. As far as image quality of CPPD-TV is concerned, the image edges and profiles were blurred with severe blocky artifacts. Finally, the regularized parameters need to be carefully tuned, which present a difficulty in practice.

To demonstrate advantages of the proposed SUGAR network over other state-of-the-art reconstruction networks, the LEARN and FISTA networks reconstruction results were also given in Figure 3 for case 1. LEARN oversmoothed image details and edges. Compared with LEARN and CPPD-TV, the newly developed FISTA method produced better images with shaper edges and finer features. However, there are still features and details missing.

Relative to the compared networks and compressed sensing results, our SUGAR offers improved image quality with clear features by unrolling the selected iterative reconstruction scheme. Specifically, the image structure indicated by the red arrow was well preserved in our SUGAR results, while it is hard to see in the CPPD-TV reconstruction. This image structure was not visualized well in the LEARN results, and not distinguishable either in the FISTA results. Fortunately, our proposed SUGAR method recovered image features significantly better, which was further confirmed by the details indicated by the green arrow. In reference to the image structures indicated by the yellow arrow, the results reconstructed using all competing methods were corrupted and distorted. This is because the measurements coming from a challenging imaging geometry limited by a short scanning arc of only 151.875°. Moreover, the image structure indicated by the arrows were severely compromised by limited-angle artifacts. In contrast, our proposed SUGAR results provided clearer image edges and structures. To evaluate the SUGAR reconstruction results volumetrically, representative sagittal and coronal slices (cases 2 and 3) are presented in Figures 4 and 5 respectively. These cases show consistently that the imaging performance of our proposed SUGAR network in terms of structural features is better than compressed sensing based reconstruction and other deep reconstruction networks for ultra-low-dose CT imaging.

Furthermore, compared with the compressed sensing based construction methods, the deep learning reconstruction methods run much faster. For example, to reconstruct an image, LEARN, FISTA and SUGAR only consumed 0.159, 0.130, 0.555 seconds, respectively.

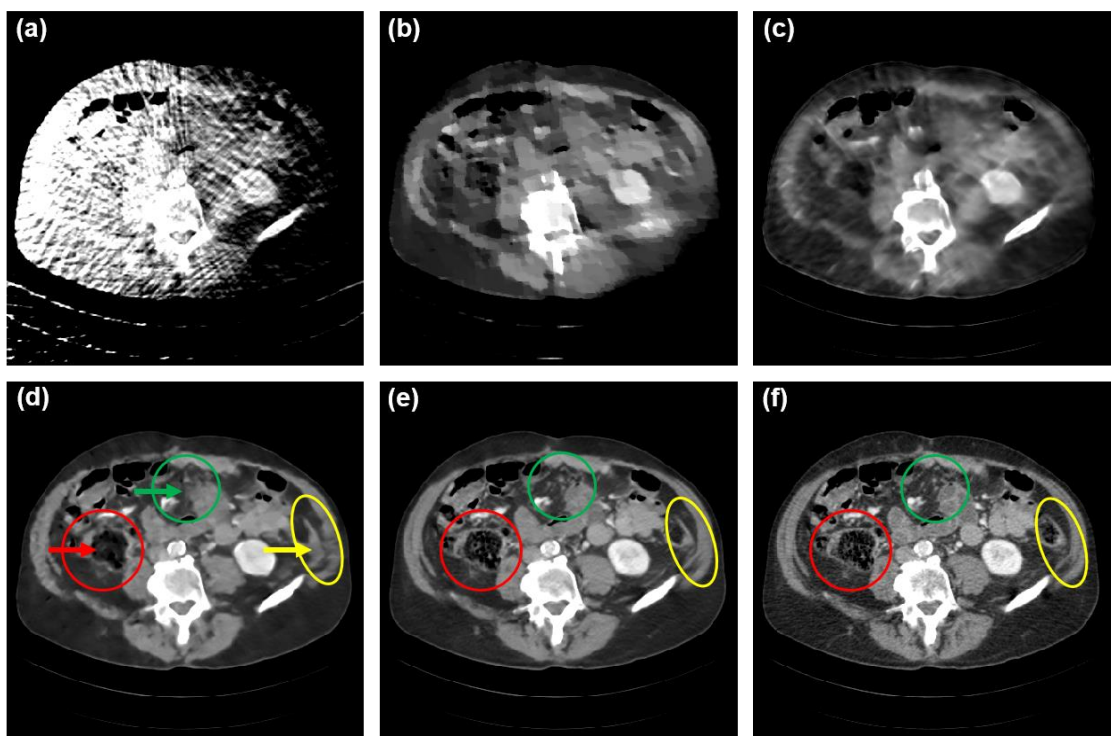

Figure 3. Transverse abdominal CT images in the same patient reconstructed from 36 projections (a-e) and the full scan (f). The images in (a-e) were reconstructed using FBP, CPPD-TV algorithm, LEARN, FISTA and our SUGAR network, respectively. The images in (a-c) have unacceptable image artifacts. The artifacts in (d) mess up image structures. The small features and image edges are seen only on our SUGAR image (e) in reference to the full-scan reconstruction (f). The display window is [-160 240] HU.

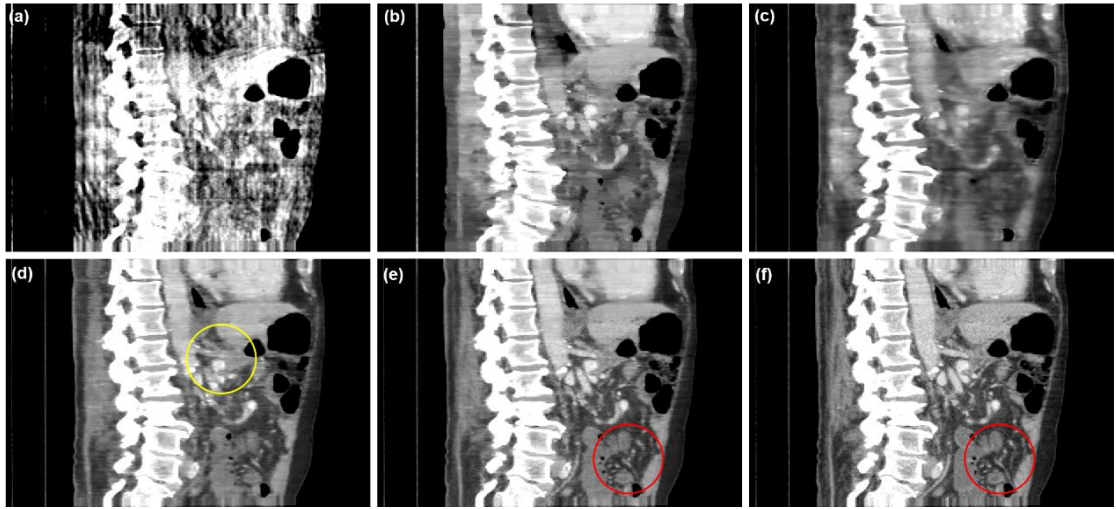

Figure 4. Sagittal abdominal CT images of the same patient as in Figure 3 with full projections (f) from 36-projection and (a)-(e) reconstructed with different reconstruction techniques. Only SUGAR-enabled image (e) demonstrated acceptable image quality from perspective of image texture, visualization of small structures (red circle), and lack of major artifacts. Other images (a)-(d) had unacceptable image artifacts, texture and/or visibility of small structures.

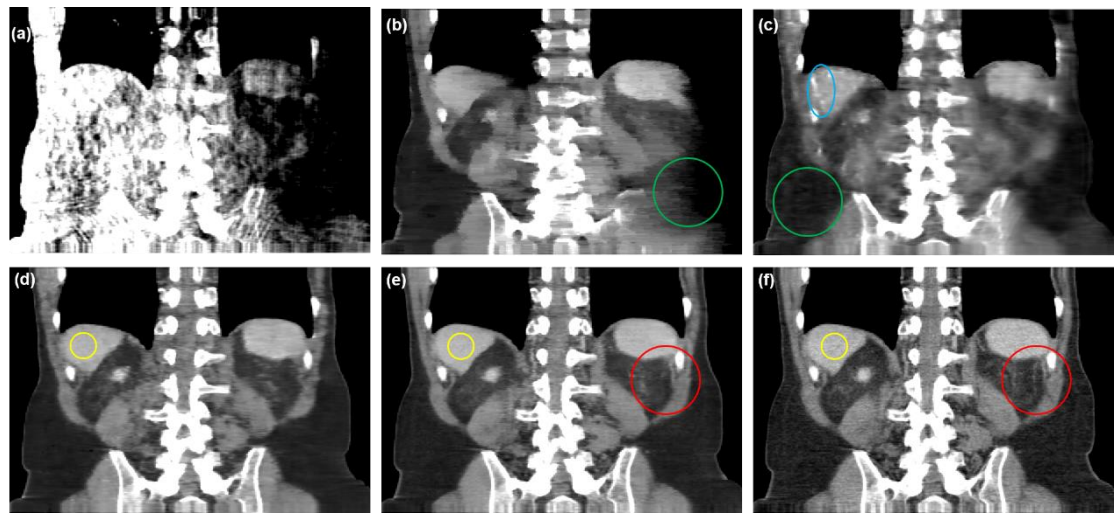

Figure 5. Coronal images of abdominal CT in the same patient as in Figure 3. The images (a-e) from sparse data using different methods demonstrate remarkable image quality variations. The images (a-c) are unacceptable for strong artifacts, poor texture and inability to assess small and/or large structures. The image (d) has little artifacts but is too smooth with unacceptable texture for clinical usability. The image (e) reconstructed with our SUGAR network has optimal image quality in terms of texture (yellow circle), no major artifacts, and acceptable visualization of small structures (red circle).

**(2) SUGAR Advantages in Terms of Lesion Detection**

To further demonstrate the advantages of SUGAR in terms of lesion detection, three orthogonal views from the patient L291 were analyzed. The transverse case is shown in Figure 6, and it is observed that our proposed SUGAR provided the best image features as well as clearest image edges. On the other hand, the lesion was completely distorted in the FBP, CPPD-TV and LEARN results. Although the lesion was, to some extent, observed in the FISTA results, the image edges are not clear. In contrast, the lesion is clearly observed in our proposed SUGAR results. Additionally, the reconstruction results in the sagittal and coronal views further demonstrate that our SUGAR method enables accurate detection of the lesion, as shown in Figures 7 and 8.

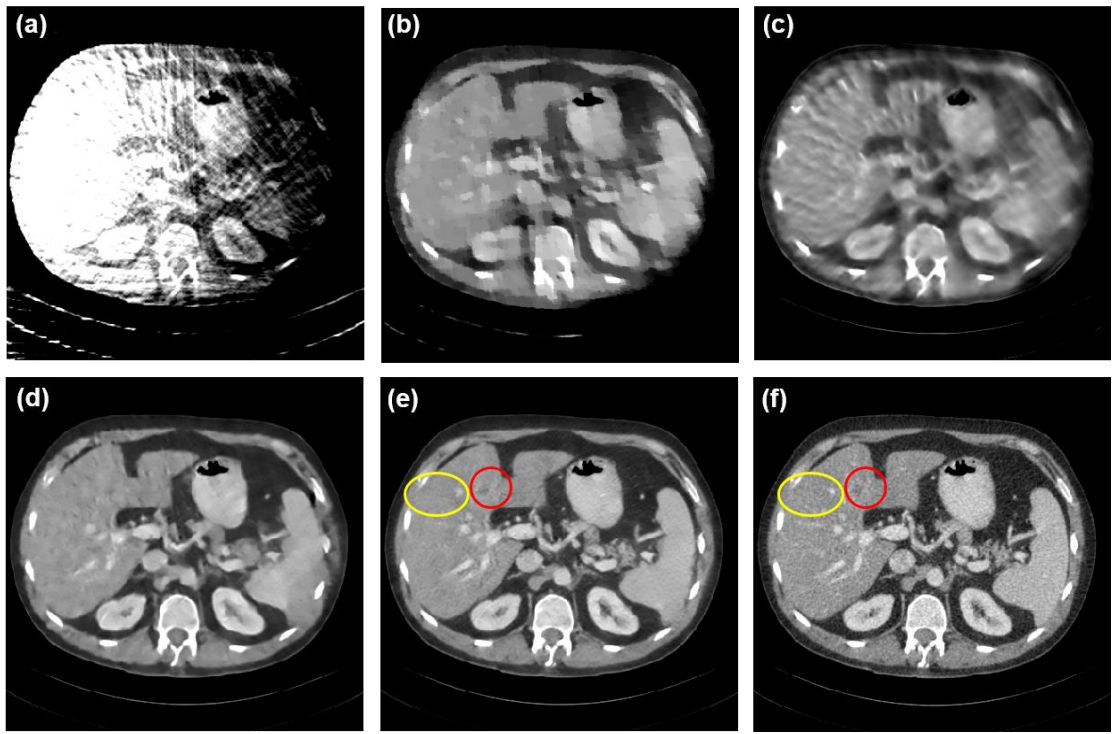

Figure 6. Transverse abdominal CT images in the same patient with full projections (a) and 36-projection data (a)-(e) reconstructed with different reconstruction. Sparse projection images (a)-(e) have unacceptable image artifacts, texture and/or visibility of small structures. The small, low contrast lesion in the left hepatic lobe lesion is seen only on acceptable full projection (f) and reduced-projection SUGAR (e) images. Note that the liver lesion is not visualized on other images including image D which has poor image texture and artifacts in the liver parenchyma.

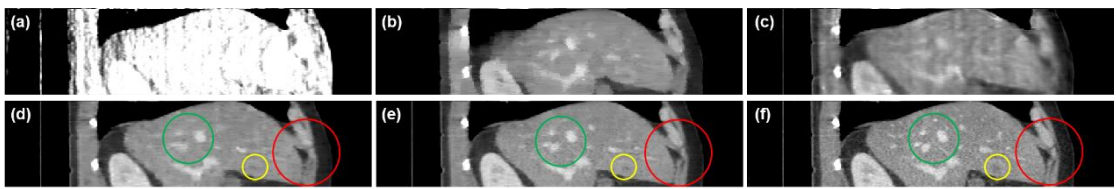

Figure 7. Sagittal abdominal CT images of the same patient as shown in Figure 6. Only our SUGAR-enabled image (e) demonstrated acceptable image quality from perspective of image texture, visualization of small structures (yellow circle), and lack of major artifacts. The images (a-c) have unacceptable artifacts, compromised texture and/or visibility of small structures, and hepatic lobe lesion cannot be observed at all. The image (d) has better quality than (a-c) but it contains shallow artifacts (red circle), blurring image structures (yellow circle) and rendering the lesion difficult to be discriminated.

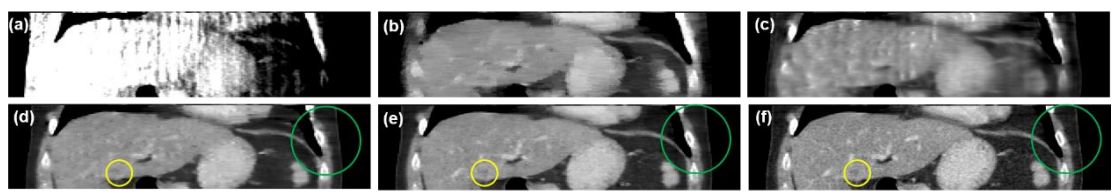

Figure 8. Coronal images of abdominal CT in the same patient as shown in Figure 6. The images (a-c) are deemed unacceptable for artifacts (green circles), poor image texture and inability to assess small and/or large structures. The image (d) has shallow (blue circle) artifacts and mess up the hepatic lobe lesion. The image (e), reconstructed with our SUGAR network, has optimal image quality from perspective of decent image texture (yellow circle), no major artifacts, and acceptable visualization of small structures.

**(3) SUGAR Advantages in Terms of Quantitative Evaluation**

To compare the results of different reconstruction methods quantitatively, the metric results in the above representative six slices are summarized in Table 1. These metrics are PSNR and SSIM. The data show that our SUGAR always achieves the highest PSNRs and SSIMs in all cases. In other words, our SUGAR results are always better than CPPD-TV, LEARN and FISTA results.

Table I. Quantitative results from all the reconstruction methods in terms of PSNR and SSIM.

|  | Cases | 1 | 2 | 3 | 4 | 5 | 6 |
|---|---|---|---|---|---|---|---|
| PSNR(dB) | FBP | 20.06 | 29.26 | 20.42 | 17.86 | 17.18 | 16.03 |
|  | CPPD-TV | 29.71 | 32.32 | 25.25 | 28.06 | 31.87 | 28.01 |
|  | LEARN | 35.86 | 35.77 | 31.89 | 31.94 | 33.86 | 29.83 |
|  | FISTA | 41.32 | 41.13 | 37.95 | 39.21 | 38.97 | 34.77 |
|  | SUGAR | 45.19 | 45.21 | 42.27 | 42.14 | 41.80 | 39.47 |
| SSIM | FBP | 0.4453 | 0.6220 | 0.4937 | 0.4416 | 0.3480 | 0.4234 |
|  | CPPD-TV | 0.8564 | 0.8379 | 0.7553 | 0.8326 | 0.8439 | 0.8000 |
|  | LEARN | 0.9004 | 0.8589 | 0.8180 | 0.8532 | 0.8261 | 0.7714 |
|  | FISTA | 0.9605 | 0.9287 | 0.9180 | 0.9451 | 0.8987 | 0.8529 |
|  | SUGAR | 0.9813 | 0.9635 | 0.9541 | 0.9685 | 0.9343 | 0.9108 |

**(4) Ablation Study**

Because the whole reconstruction process was implemented in two steps: LE and HR. A natural question is about the performance of the HR step. To validate the performance, both LE and HR difference images in two cases are given in Figure 9. It is seen that the HR step improves the reconstruction performance with error reduction. Furthermore, we evaluated the performance of directly reconstructing high-dimensional images (i.e., 512x512 pixels) from limited data. For that purpose, we repeated the reconstruction process to output high-dimension images without utilizing intermediate reconstructions (256x256 pixels). The reconstruction results with direct reconstruction in case 1 are given in Figure 10. It is clear that the direct reconstruction degraded the image quality significantly. Indeed, our proposed two-stage reconstruction strategy allows much clearer image edges and other features.

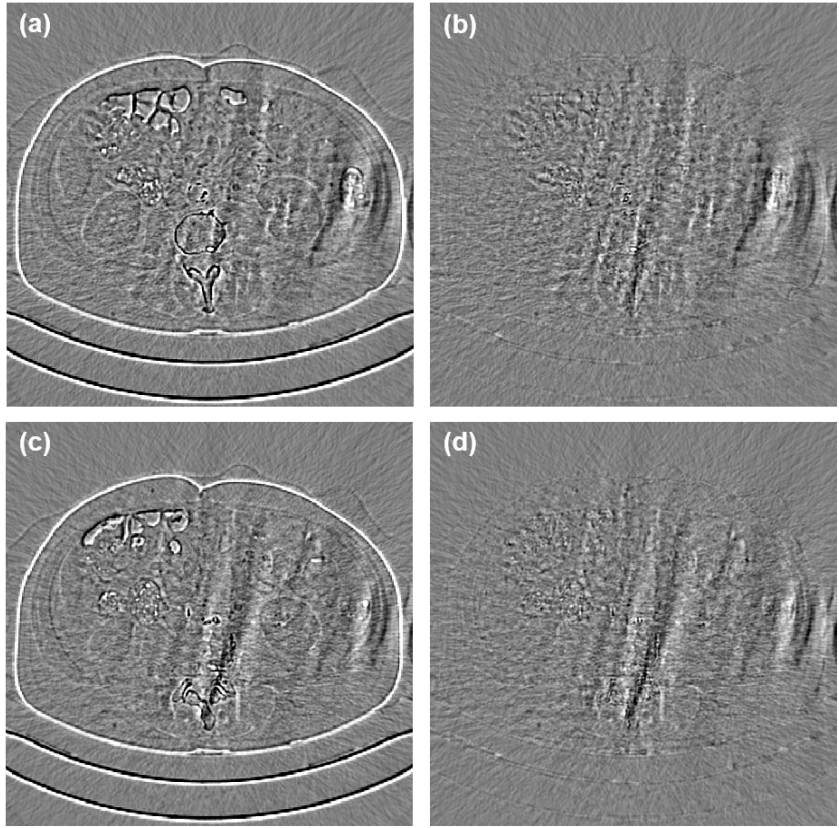

Figure 9. Comparison results with/without the high-dimensional refinement in both two cases. The 1st and 2nd rows represent the two cases respectively. The left and right column images are the reconstructed results without and with high-dimensional refinement. The display window is [-80 80] HU.

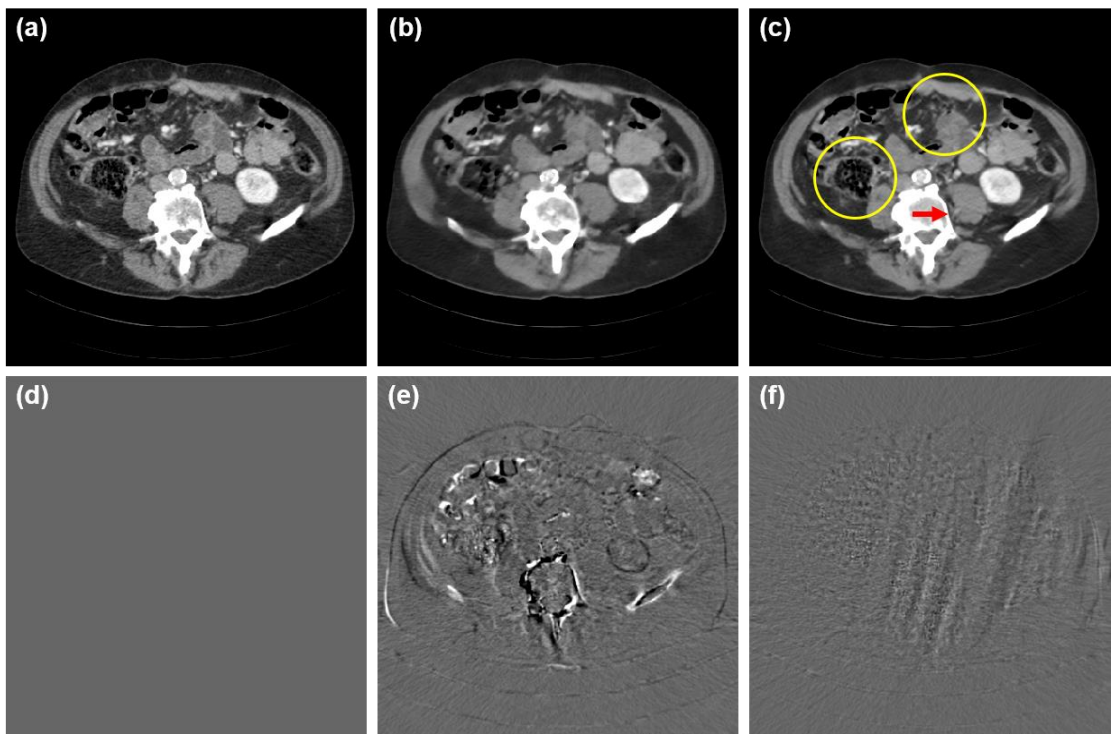

Figure 10. Comparison between the direct reconstruction into a 512x512 image from measurement and the staged reconstruction as illustrated in Figure 2(a). The 1st and 2nd rows represent image and the corresponding residual

respectively, while the 1st to 3rd columns are the ground truth, direct and staged reconstructions respectively. The display window is [-80 80] HU.

**IV. Discussions & Conclusion**

Based on the clinical results from our experiments, our deep learning based SUGAR method can achieve high-quality images. Specifically, there are several key features and merits of SUGAR. First, we have unified the split-Bregman iteration and data-driven deep network-based reconstruction. Second, to fully explore deep priors from big data and optimize the regularization parameters, we have formulated a non-linear sparsifying function in terms of a neural network and incorporated it into a general reconstruction model to solve the sparse-data reconstruction problem effectively and efficiently. Third, to meet the challenge of high-resolution reconstruction using the unrolling-based deep reconstruction strategy, our strategy can recover high resolution CT images in two steps: LE and HR.

Our network has been shown successful in improving the reconstruction performance significantly. Compared with the CS-based reconstruction methods (CPPD-TV), the advantages of SUGAR are prominent in the following aspects: (1) removing the burden of the selection of parameters in specific applications; (2) reducing the computational cost for fast imaging; and (3) achieving a major reconstruction quality gain over existing methods for sparse-data CT reconstruction. In particular, compared with recent deeply unrolled tomographic networks, including LEARN [33] and FISTA-net [38], the merits of our SUGAR are demonstrated in the following aspects: (1) SUGAR is an interpretable deep tomographic network, which is motivated by a rigorous derivation based on the split-Bregman optimization; (2) the encoder-decoder neural block is employed to facilitate transforms between data and image domains, where the sampling processes are implemented as multiple-level down-sampling convolutional layers for feature extraction and up-sampling convolutional operators for image reconstruction; and (3) our SUGAR reconstruction works in two-steps for low-dimensional and high-dimensional reconstruction.

Future studies with a larger cohort of patients will be needed to establish how SUGAR compares with current filtered backprojection and other deep learning-based image reconstruction techniques with full views/projections for image reconstruction in patients scanned under ULD-CT protocols. An important observation in our feasibility study is the retained image texture and conspicuity of subtle low-contrast lesions in the SUGAR reconstructed ULD-CT images [44]. If these advantages of SUGAR hold up in a larger patient dataset, it could lead to wider adoption of ULD-CT in clinical practice, especially with a potential to acquire sparsely-sampled data.

In conclusion, we have explored the feasibility of ultra-low-dose CT in clinical applications using numerical optimization and deep learning technologies by synergizing deep prior with an iterative reconstruction framework into a unified neural network referred to as the SUGAR network. SUGAR has consistently produced reconstruction results superior to the state-of-the-art compressed sensing based reconstruction and existing unrolling-based reconstruction networks. It is emphasized that our SUGAR network serves as an example to show the power of deep learning based reconstruction methods in the ultra-low-dose setting. We believe that deep learning has a huge potential in tomographic imaging for ultra-low-dose CT applications.